\documentclass[10pt,twocolumn,letterpaper]{article}

\usepackage{cvpr}
\usepackage{times}
\usepackage{epsfig}
\usepackage{graphicx}
\usepackage{amsmath}
\usepackage{amssymb}
\usepackage[utf8]{inputenc} 
\usepackage[T1]{fontenc}    
\usepackage{url}            
\usepackage{booktabs}       
\usepackage{amsfonts}       
\usepackage{nicefrac}       
\usepackage{microtype}      
\usepackage{amsmath}
\usepackage{enumitem}
\usepackage[numbers,sort&compress]{natbib}
\usepackage{xcolor}
\usepackage{float}

\setlist[enumerate]{itemsep=0mm}

\newcommand{\x}{\mathbf{x}}

\newcommand{\K}{\mathbf{k}}

\newcommand{\ft}{\mathcal{F}}
\newcommand{\ift}{\mathcal{F}^{-1}}

\newcommand{\C}{\mathbb{C}}

\newcommand{\norm}[1]{\left\lVert#1\right\rVert}
\newcommand{\abs}[1]{\left\lvert#1\right\rvert}

\DeclareMathOperator{\argmin}{argmin}
\DeclareMathOperator{\ssim}{SSIM}

\usepackage[pagebackref=true,breaklinks=true,letterpaper=true,colorlinks,bookmarks=false]{hyperref}

\cvprfinalcopy 

\def\cvprPaperID{9974} 

\ifcvprfinal\fi

\title{GrappaNet: Combining Parallel Imaging with Deep Learning for Multi-Coil MRI Reconstruction}

\author{
Anuroop Sriram$^1$ \quad\quad Jure Zbontar$^1$ \quad\quad Tullie Murrell$^1$ \quad\quad C. Lawrence Zitnick$^1$\\
Aaron Defazio$^1$ \quad\quad\quad Daniel K. Sodickson $^2$ \vspace{.5em} \\
$^1$Facebook AI Research (FAIR)  \quad $^2$NYU School of Medicine\vspace{-.5em}
}

\begin{document}
\maketitle
\thispagestyle{empty}

\maketitle

\begin{abstract}

Magnetic Resonance Image (MRI) acquisition is an inherently slow process which has spurred the development of two different acceleration methods: acquiring multiple correlated samples simultaneously (parallel imaging) and acquiring fewer samples than necessary for traditional signal processing methods (compressed sensing). Both methods provide complementary approaches to accelerating MRI acquisition. 

In this paper, we present a novel method to integrate traditional parallel imaging methods into deep neural networks that is able to generate high quality reconstructions even for high acceleration factors. The proposed method, called GrappaNet, performs progressive reconstruction by first mapping the reconstruction problem to a simpler one that can be solved by a traditional parallel imaging methods using a neural network, followed by an application of a parallel imaging method, and finally fine-tuning the output with another neural network. The entire network can be trained end-to-end. We present experimental results on the recently released fastMRI dataset \cite{zbontar2018fastmri} and show that GrappaNet can generate higher quality reconstructions than competing methods for both $4\times$ and $8\times$ acceleration.

\end{abstract}

\section{Introduction}

Magnetic Resonance Imaging (MRI) is the leading diagnostic modality for a wide range of disorders including musculoskeletal, neurological,
and oncological diseases. However, the physics of the MRI data acquisition process make it inherently slower than alternate modalities like CT or X-Ray. As a consequence, increasing the speed of MRI acquisition has been a major ongoing research goal for decades.

Parallel Imaging (PI) is one of the most important and successful developments in reducing MRI scan time~\cite{pruessmann1999sense,griswold2002Grappa}. The technique requires the use of multiple physical receiver coils to simultaneously record different views of the object being imaged. Parallel imaging is the default option for many scan protocols and it is supported by almost all modern clinical MRI scanners.

Another approach to accelerating MR imaging is the use of Compressed Sensing (CS), which can speed up MRI acquisition by acquiring fewer samples than required by traditional signal processing methods. To overcome aliasing artifacts introduced by violating the Shannon-Nyquist sampling theorem, CS methods incorporate additional a priori knowledge about the images. Recently, the use of learned image priors through the use of deep learning have rapidly gained in popularity~\citep{varnet,WangSYPZLFL16,SchlemperCHPR17,Zhu2018}. These approaches have shown a significant improvement in image reconstruction quality, particularly for non-parallel MRIs.

In this paper, we show that a novel combination of classical parallel imaging techniques with deep neural networks can achieve higher acceleration factors than using either approach alone. Utilizing parallel imaging in deep learning approaches to reconstruction is challenging. The relation between the captured views changes for each scan and is dependent on the configuration of the detectors with respect to the object being imaged.

To address this challenge we introduce GrappaNet, a new neural network architecture that incorporates parallel imaging. GrappaNet contains a GRAPPA layer that learns a scan-specific reconstruction function to combine the views captured during parallel imaging. To allow the network to fully utilize all the information captured during parallel imaging, the reconstruction is performed jointly across all the complex-valued views captured during the parallel imaging process. Unlike many previous approaches~\cite{varnet}, the views are not combined until the final layer to produce the output reconstruction. The model uses a progressive refinement approach in both k-space (frequency domain) and image space to both aid in the optimization and to take advantage of the complementary properties of the two spaces. Most previous approaches typically focus on either reconstructing in image space~\cite{varnet} or k-space~\cite{HanYY17}. We evaluate the performance of our method on the recently released fastMRI \cite{zbontar2018fastmri} dataset.

First, we present a brief introduction to parallel MR imaging and review some deep learning methods for parallel MRI reconstruction in section \ref{sec:background}. Next, we provide a description of the GrappaNet model in section \ref{sec:grappa_net}, followed by a description of our experiments in section \ref{sec:expts}. Finally, we conclude with a discussion of future work in section \ref{sec:conclusions}.

\section{Background and Related Work}\label{sec:background}

\subsection{Parallel MRI}\label{sec:parallel_mri}
MR scanners image a patient's anatomy by acquiring measurements in the frequency domain using a measuring instrument called a receiver coil. In the MRI literature, these frequency-domain measurements are called \emph{k-space} samples, where k refers to the spatial wave number. The image can then be obtained by applying an inverse multidimensional Fourier transform $\ift$ to the measured k-space samples.
The underlying image $\x \in \C^M$ is related to the measured k-space samples $\K \in \C^M$ as
\begin{equation}
    \K = \ft(\x) + \epsilon,
\end{equation}
where $\epsilon$ is the measurement noise.

Most modern scanners support parallel imaging: they employ an array of multiple receiver coils that simultaneously obtain k-space samples from the anatomy being imaged. The k-space samples measured by each coil are modulated by their sensitivity to the MR signal arising from different regions. In particular, the k-space sample measured by the $i$-th coil is
\begin{equation}
    \K_i = \ft(S_i\x) + \epsilon_i, i = 1, 2, \dots, N,
\end{equation}
where $S_i$ is a complex-valued diagonal matrix encoding the position dependent sensitivity map of the $i$-th coil and $N$ is the number of coils.

Different coils are typically sensitive to different but overlapping regions. It is important to note that the coil sensitivities vary per scan since they depend not only on the configuration of the coils but also on their interaction with the anatomy being imaged.

\subsection{Accelerated MRI}\label{sec:accel_mri}
The speed of MRI acquisition is limited by the number of k-space samples obtained. This process can be accelerated by obtaining only a subset of the k-space data:
\begin{equation}
    \K_i = M \ft(S_i\x) + \epsilon_i, i = 1, 2, \dots, N,
\end{equation}
where $M$ is a binary mask operator that selects a subset of the k-space points. The same mask is used for all coils. Applying an inverse Fourier transform naively to this under-sampled k-space data results in aliasing artifacts.

Parallel MRI can be used to accelerate imaging by exploiting the redundancies in k-space samples measured by different coils to estimate the missing k-space points from the observed points. Various parallel imaging methods have been proposed but they can be divided into two broad classes: a) \emph{SENSE}-type methods \cite{pruessmann1999sense} that operate in the image space, and b) \emph{GRAPPA}-type methods \cite{griswold2002Grappa} that operate locally in k-space. The latter is relevant to this work.

The GRAPPA algorithm estimates the unobserved k-space points as a linear combination of the neighboring observed k-space points from all coils. The same set of weights are used at all spatial locations, which can be seen as a complex-valued convolution in k-space from $N$ channels to $N$ channels, where $N$ is the number of coils. Formally, the unobserved k-space points $\K^u$ are computed from the observed k-space points $\K$ by convolving with GRAPPA weights $G$:
\begin{equation} \label{eqn:grappa}
    \K^u = G * \K.
\end{equation}

During acquisition, the central region of k-space (which corresponds to low spatial frequencies) is fully sampled. This region, called the Auto-Calibration Signal or ACS, is used to estimate the GRAPPA weights $G$. We can simulate under-sampling in the ACS by masking out certain k-space points. 
Let the simulated observed and unobserved k-space points in the ACS be $\K'$ and $\K^{u\prime}$ respectively. 
From equation \ref{eqn:grappa}, the convolution of $G$ and $\K'$ should be equal to $\K^{u\prime}$. 
Thus, we can estimate $G$ by solving the following optimization problem:
\begin{equation} \label{eqn:grappa_calib}
    \hat{G} = \argmin_G \norm{\K^{u\prime} - G * \K'}^2.
\end{equation}

The knee images in the fastMRI dataset~\cite{zbontar2018fastmri} were acquired using machines that employ 15 receiver coils and can generally support 2$\times$ acceleration for imaging of the knee using this approach. Higher acceleration factors lead to aliasing artifacts that cannot be removed by standard parallel imaging methods.

\subsection{Compressed Sensing for Parallel MRI Reconstruction}\label{sec:cs_mri}

Compressed Sensing \cite{donoho2006compressed} enables reconstruction of images by using fewer k-space measurements than is possible with classical signal processing methods by enforcing suitable priors. Compressed sensing has been combined with parallel imaging to achieve higher acceleration factors than those allowed by parallel imaging alone.

Classical compressed sensing methods use sparsity in some transform domain as a prior. Many classical compressed sensing methods operate in the image domain and solve the following optimization problem:
\begin{equation}\label{eq:pics_mri}
    \hat{\x} = \argmin_{\x} \frac{1}{2} \sum_{i} \norm{\ft(S_i \x) - \K_i}^2 + \lambda \Psi(\x),
\end{equation}
where $\Psi$ is a regularization function that enforces a sparsity constraint in some transform domain such as gradients in the image domain. This problem can be solved by iterative gradient descent style methods.

In the last few years, there has been rapid development of deep learning based approaches to MRI reconstruction~\cite{WangSYPZLFL16,SchlemperCHPR17,Zhu2018,liang2019deep,jin2016inv,varnet}.
MRI reconstruction can be viewed as an inverse problem and several previous research papers have proposed neural networks whose design is inspired by the optimization procedure to solve such an inverse problem~\cite{liang2019deep,jin2016inv,varnet}.
One approach in this direction for the multi-coil reconstruction problem is the Variational Network (VN) ~\cite{varnet}. The VN model is a deep neural network, each of whose layers implements one gradient update step for the optimization problem in equation \ref{eq:pics_mri}. The VN uses pre-computed sensitivity maps and achieves excellent reconstructions at low acceleration factors. Computing sensitivity maps becomes more challenging at higher accelerations, which may limit the maximum acceleration this method can achieve.

An alternate line of work operating in k-space is the RAKI model \cite{akcakaya2019raki} which replaces the single convolution operation in GRAPPA with a deep convolutional network that is trained independently for each scan. The RAKI method emphasizes the importance of using a scan specific model for multi-coil reconstruction. This method is complementary to our work and can be integrated into the GrappaNet by replacing the GRAPPA layer with the RAKI network.

A comprehensive survey of recent developments in using deep learning for parallel MRI reconstruction can be found in \cite{Knoll2019DeepLM}.

\section{GrappaNet}\label{sec:grappa_net}

\begin{figure*}
\begin{center}
\includegraphics[width=\textwidth,keepaspectratio]
{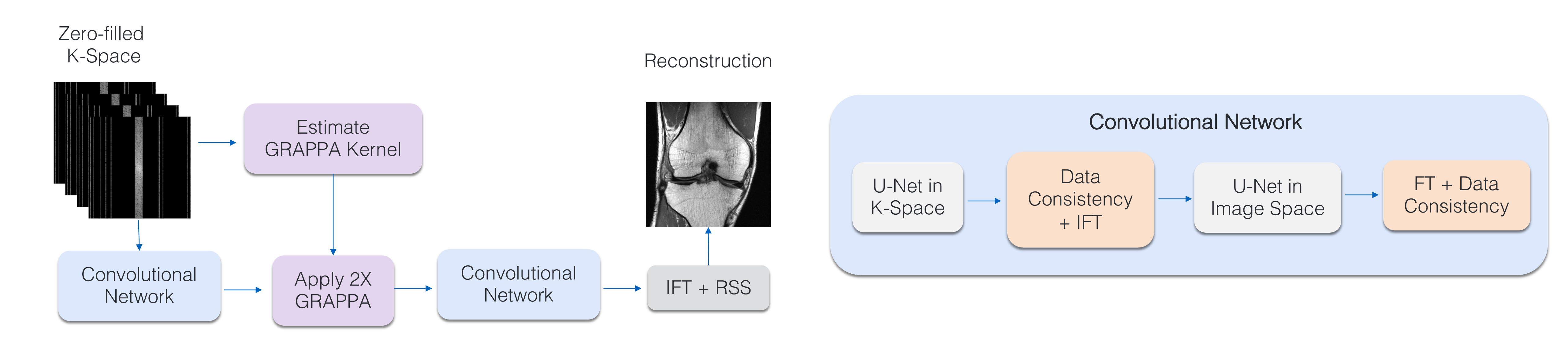}
\caption{\label{fig:grappanet_model} Left: The full GrappaNet model which takes under-sampled k-space samples as input and outputs the reconstructed image. Right: Details about each of the convolutional networks, which take multi-coil k-space as input and output multi-coil k-space. Here, FT, IFT \& RSS refer to 2D Fourier transform, 2D inverse Fourier transform and root sum-of-squares operations (equation \ref{eq:rss}) respectively.}
\end{center}
\end{figure*}

The GrappaNet is a neural network that takes under-sampled, multi-coil k-space data as input and outputs the reconstructed image. Figure \ref{fig:grappanet_model} shows a diagram of the network architecture that contains three important properties. First, the differentiable GRAPPA layer enables the network to take advantage of the known physical proprieties of parallel imaging. Next, each convolutional network is applied across all complex-valued views jointly, before being combined in the final stage. This enables the network to take advantage of all the information captured during parallel imaging. Several previous approaches \cite{varnet, HanYY17}, performed reconstruction after collapsing to a single view.
Finally, image-to-image mappings using U-Nets are performed in both k-space and image space. Convolutions, pooling, and up-sampling result in very different operations in image space and k-space. We demonstrate in Section~\ref{sec:expts} that using both these complementary spaces provides improved accuracy.

The network consists of two convolutional neural networks, with the application of the GRAPPA operator in between them. Denoting the input under-sampled k-space data by $\K$, the network computes the following function: 
\begin{equation}
{\x} = h \circ f_2(G * f_1(\K)),
\end{equation}
where $f_1$, and $f_2$ are convolutional networks that map multi-coil k-space to multi-coil k-space and $h$ combines the multi-coil k-space data to a single image by first applying an inverse fourier transform followed by an root sum-of-squares (RSS) operation (equation \ref{eq:rss}).

The first network, $f_1$ takes the multi-coil k-space data with $R$-fold under-sampling and maps it to an $R'$-fold under-sampled k-space dataset with the same number of coils. The GRAPPA operator, $G$, which is separately obtained from the ACS, is then applied to this $R'$-fold under-sampled dataset to fill in the missing k-space data. This allows the network to take advantage of the known physical proprieties of the parallel imaging process. $R'$ is chosen to be small enough that traditional parallel imaging methods like GRAPPA can reconstruct the image accurately. We use $R'=2$ for our experiments.

\subsection{U-Net}

Both $f_1$ and $f_2$ are composed of multiple U-Nets ~\cite{Ronneberger2015unet}, which are convolutional networks that operate at multiple scales. U-Net models and their variants have successfully been used for many image-to-image mapping tasks including MRI reconstruction~\cite{Hyun2018DeepLF,Han2017} and image segmentation~\cite{DBLP:journals/corr/RonnebergerFB15}. The U-Nets used in this work are based on the U-Net baseline models from ~\cite{zbontar2018fastmri}.



A U-Net is useful for image to image mapping tasks like semantic segmentation because the presence of pooling and up-sampling layers allow it to learn useful feature maps at multiple scales and abstraction levels. This multi-resolution feature representation helps the U-Net predict the higher level details of the output at the lowest level of the decoder and gradually adds finer, higher frequency details as the up-sampling layers are applied.

The baseline model, described in~\cite{zbontar2018fastmri}, used such a U-Net model for MRI image reconstruction. However, that model is only able to perform denoising since it is applied after combining the different views using a root-sum-of-squares (RSS) transform (equation \ref{eq:rss}). This prevents the baseline model from learning how to combine all of the coils and using the phase information. As a result, the reconstructions from this baseline model are too smooth and lose much of the medically relevant high frequency information (see Figure \ref{fig:example-8x}). We show in section~\ref{sec:expts} that simply applying a U-Net to the real and imaginary data from all coils can significantly improve upon this model. Such a U-Net can potentially learn to combine information from different coils together, which improves performance.

Han et al \cite{HanYY17} show that a U-Net can also be applied directly to under-sampled k-space data. Their work was motivated by connections between encoder-decoder models and a classical CS algorithm called the annihilating filter-based low-rank Hankel matrix approach (ALOHA)~\cite{jin2015aloha}. The input to the ALOHA U-Net is zero-filled k-space data and the model fills in the missing information. In an approach similar to the fastMRI baseline model~\cite{zbontar2018fastmri}, Han et al~\cite{HanYY17} also apply their U-Net after combining all of the coils into a single coil. Taking insight from algorithms like GRAPPA, we posit that it would be beneficial to apply convolutions directly to the multi-coil k-space data. We show in section \ref{sec:expts} that such a model outperforms the baseline models.

 The functions $f_1$ and $f_2$ apply the following series of operations to the input k-space data (see figure \ref{fig:grappanet_model}): a U-Net in k-space followed by a hard data consistency, inverse 2D Fourier transform to convert to image space, a U-Net in the image space, followed by a 2D Fourier transform and data consistency. Each of the U-Nets map 15 complex-valued channels to 15 complex-valued channels. Here, the hard data consistency operations simply copy all of the observed k-space samples to the correct locations in k-space. This ensures that the model only fills in the missing k-space points.

The function $h$ combines the reconstructed multi-coil k-space data into a single real-valued image by first applying an inverse 2D Fourier transform to each coil, followed by a root sum-of-squares (RSS) operation. The RSS operation combines all the coils into a single real-valued image:
\begin{equation}\label{eq:rss}
    RSS(\x_1, \dots \x_N) = \left(\sum_{n=1}^{N} \abs{\x_n}^2 \right)^{1/2},
\end{equation}
where $\x_1, \dots, \x_N$ are the images from the $N$ coils.

\subsection{GRAPPA Layer}

As explained in the previous section convolutional networks in k-space or image space applied to all coils can, to a limited extent, learn to combine all of the coils. However, as described in \ref{sec:accel_mri}, the coil sensitivities can vary from one imaging examination to another. Traditional parallel imaging methods take this into consideration by estimating distinct sensitivity maps or GRAPPA kernels for each scan. This motivates the need to include a scan-specific component within the neural network that can adapt to differences in the sensitivity profile to improve generalization of the reconstruction model.

We achieve this adaptation by introducing a new neural network layer that we call the \emph{GRAPPA layer}. The GRAPPA layer estimates the GRAPPA kernel from the ACS region and then applies a two dimensional convolution with the estimated kernel. Because the application of GRAPPA is differentiable, the entire network can be trained in an end-to-end fashion using backpropagation.

\begin{figure*}
\includegraphics[page=2,width=\linewidth,keepaspectratio]{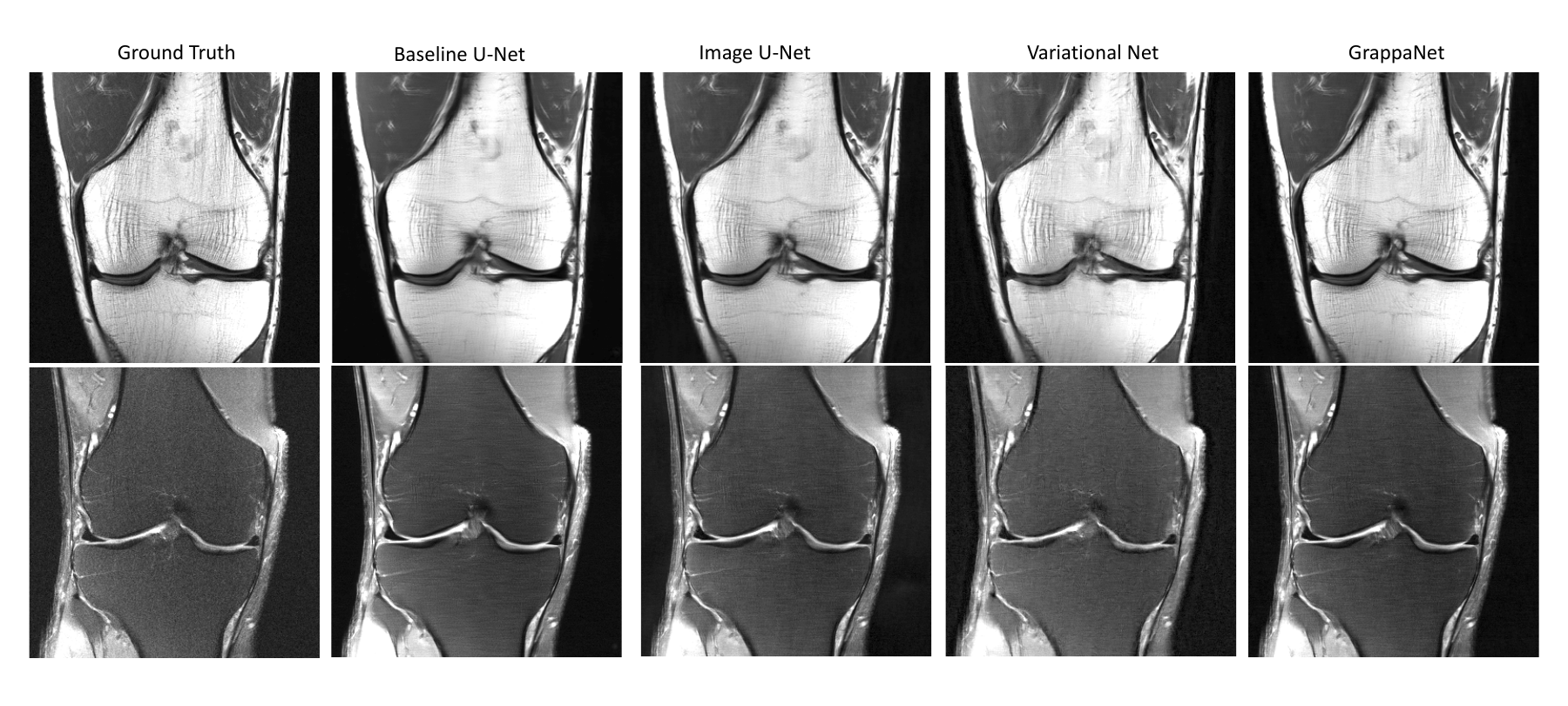}
\caption{\label{fig:example-4x} Example reconstructions for $4\times$ under-sampling. The top row shows PD images without fat suppression, and the bottom row shows PDFS images with fat suppression.}
\end{figure*}

\begin{figure*}
\centering
\includegraphics[page=4,width=\textwidth,keepaspectratio]{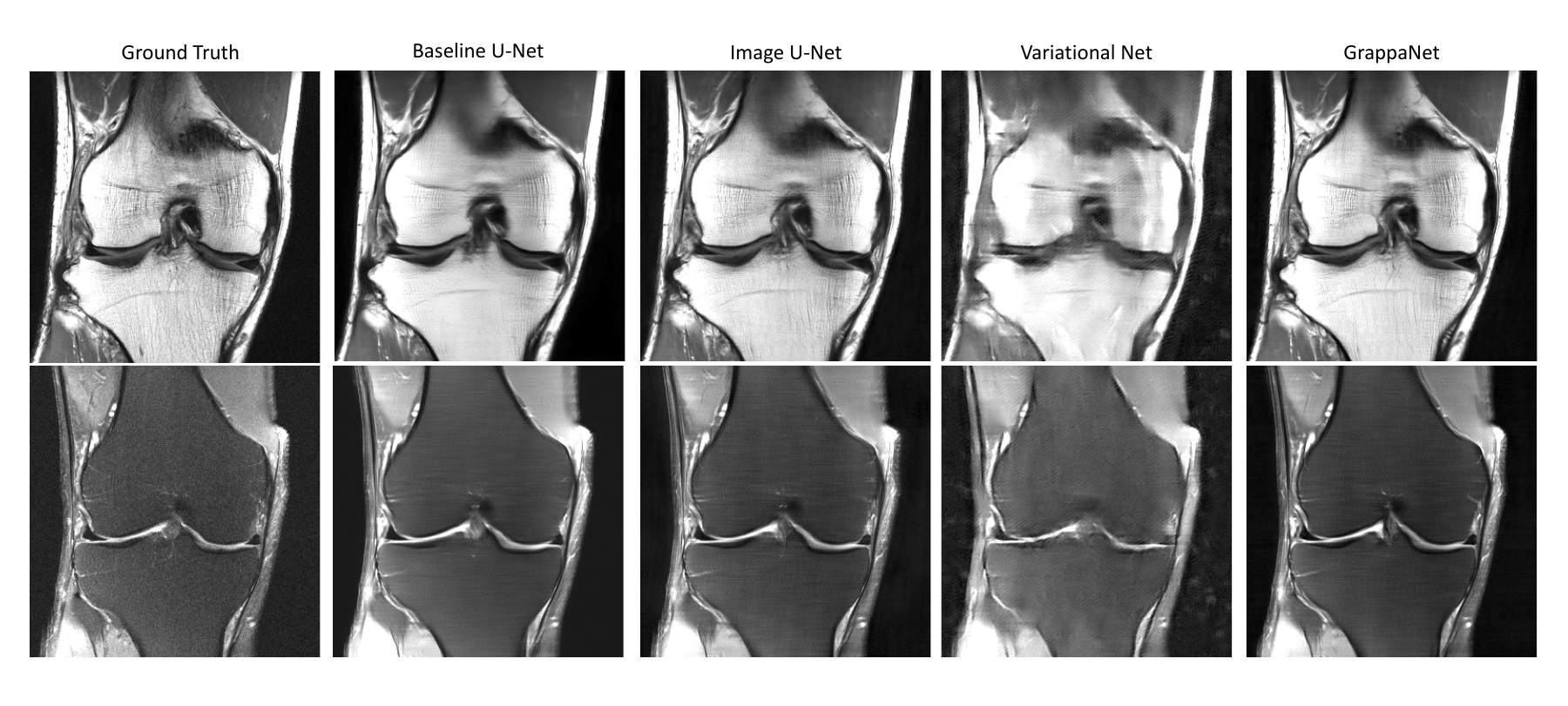}
\caption{\label{fig:example-8x} Example reconstructions for $8\times$ under-sampling. The top row shows PD images without fat suppression and the bottom row shows PDFS images with fat suppression.}
\end{figure*}

\begin{figure*}
\centering
\includegraphics[page=1,width=\textwidth,keepaspectratio]{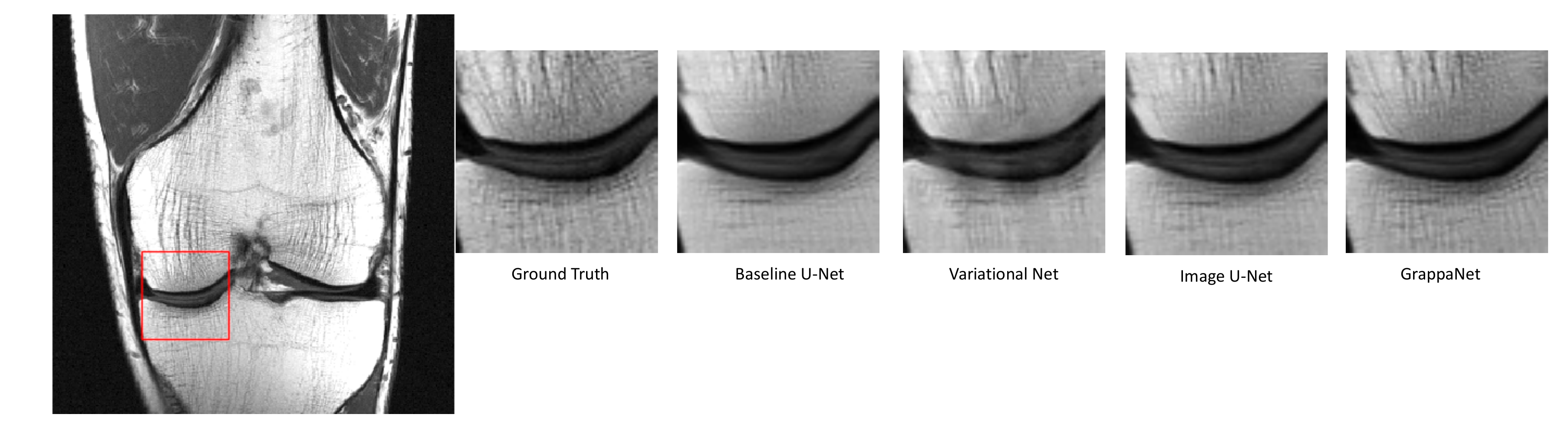}
\caption{\label{fig:example-4x-crops} Example reconstructions for $4\times$ under-sampling with the diagnostically important regions zoomed in.}
\end{figure*}

\begin{figure*}
    \centering
    \includegraphics[width=\textwidth,keepaspectratio]{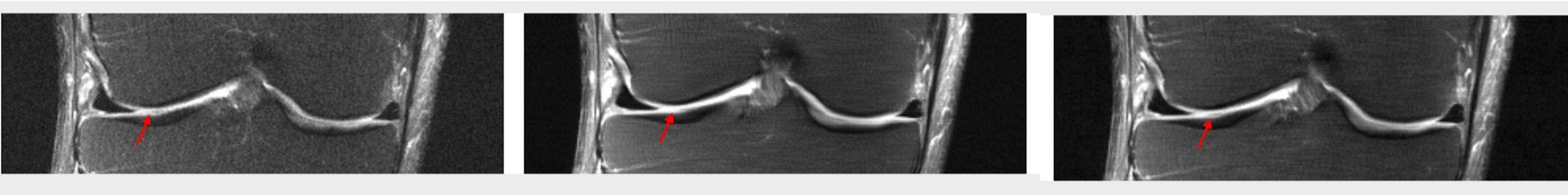}
    \caption{\label{fig:recons-arrows} Left: Ground truth, Middle: Reconstruction from baseline, Right: Reconstruction from GrappaNet. The arrow points to the meniscus region, which appears "filled in" with the baseline method. Highly accurate reconstruction of the meniscus is important for radiologists in diagnosing certain pathologies. }
\end{figure*}

\begin{table*}[htb]
\centering
\begin{tabular}{c|c|cc|cc|cc}\toprule
Acceleration & Model & \multicolumn{2}{c|}{NMSE} & \multicolumn{2}{c|}{PSNR} & \multicolumn{2}{c}{SSIM} \\ 
\cmidrule(r){3-4} \cmidrule(r){5-6} \cmidrule(r){7-8} 
& & PD & PDFS & PD & PDFS & PD & PDFS \\ \midrule

4-fold  & Classical CS baseline & 0.0198 & 0.0951 & 32.6 & 27.5 & 0.693 & 0.588 \\
        & U-Net Baseline & 0.0154 & 0.0525 & 34.00 & 29.95 & 0.815 & 0.636\\
        & Variational Net & 0.0138 & 0.0262 & 35.82 & 33.196 & 0.919 & 0.855 \\
        & K-Space U-Net & 0.0055 & 0.0114 & 37.27 & 36.45 & 0.927 & 0.870\\
        & Image U-Net & 0.0034 & 0.0103 & 39.58 & 36.97 & 0.949 & 0.886\\
        & GrappaNet & \textbf{0.0026} & \textbf{0.0085} & \textbf{40.74} & \textbf{37.77} & \textbf{0.957} & \textbf{0.891} \\
\midrule
8-fold  & Classical CS baseline & 0.0352 & 0.109 & 29.6 & 26.8 & 0.642 & 0.551\\
        & U-Net Baseline & 0.0261 & 0.0682 & 31.5 & 28.71 & 0.762 & 0.559\\
        & Variational Net & 0.0211 & 0.0816 & 32.12 & 27.72 & 0.788 & 0.675 \\
        & K-Space U-Net & 0.0189 & 0.0206 & 36.45 & 32.54 & 0.870 & 0.807\\
        & Image U-Net & 0.0079 & 0.0160 & 36.26 & 34.36 & 0.886 & 0.831\\
        & GrappaNet & \textbf{0.0071} & \textbf{0.0146} & \textbf{36.76} & \textbf{35.04} & \textbf{0.922} & \textbf{0.842}\\

\bottomrule
\end{tabular} 
\caption{Experimental results}
\label{tbl:results}
\end{table*}

\section{Experimental Results}\label{sec:expts}

We ran all our experiments on the multi-coil knee MRIs from the fastMRI dataset~\cite{zbontar2018fastmri}, which consists of raw k-space data from 1594 knee MRI exams from four different MRI machines. The dataset contains two types of MRI sequences that are commonly used for knee exams in clinical practice: a Proton Density (PD) weighted sequence and a Proton Density weighted sequence with Fat Saturation (PDFS).
We used the same train, validation and test splits as in the original dataset. The training data consisted of 973 volumes which contained k-space data of different sizes. During training, we omitted k-space data with a width greater than 372, which is about 7\% of the training data. We evaluated various models on all test images.

For training our models, we used random masks with $4\times$ and $8\times$ accelerations, based on code released with the fastMRI dataset\footnote{\label{code:fastMRI}https://github.com/facebookresearch/fastMRI}. We experimented with the following models:
\begin{enumerate}
    \item Classical CS baseline based on Total Variation minimization~\cite{zbontar2018fastmri}
    \item U-Net baseline model applied to RSS inputs~\cite{zbontar2018fastmri}
    \item Variational Network model introduced in~\cite{varnet}
    \item U-Net applied in k-space to 15 coil input
    \item U-Net applied in image space to 15 coil input
    \item GrappaNet model
\end{enumerate}

We used the original implementation of the Variational Network\footnote{https://github.com/VLOGroup/mri-variationalnetwork/}. This code runs the ESPIRiT algorithm \cite{uecker2014espirit} to estimate sensitivity maps from the densely sampled ACS region. These maps are used both as input to the network and also to combine the fully sampled coil responses to compute the training targets. For experiments with $8\times$ accelerations, the input k-space contains very few ACS lines, which yields poor quality sensitivity maps for the Variational Network. The training targets computed using these poor quality sensitivity maps contain aliasing artifacts that make them unsuitable for training. To mitigate this problem, we always use 30 low frequency lines to compute the training target for $8\times$ experiments. The sensitivity maps used as inputs to the network are still computed from the ACS region. We did not change the model architecture or training procedure from the original implementation, except for the use of random masks.

For the k-space U-Net, the image space U-Net, and the GrappaNet models, we followed the training procedure for the baseline models in \cite{zbontar2018fastmri}. To deal with complex-valued inputs, we simply treated the real and imaginary parts as two distinct channels. Hence, 15-coil complex-valued k-space or image data were treated as 30-channel data. These models were trained using the RMSProp~\cite{rmsprop} algorithm to minimize a linear combination of Structural Similarity (SSIM)~\cite{wang2004image} and $L1$ losses:
\begin{equation}
    J(\hat{\x}, \x) = -\ssim(\hat{\x}, \x) + \lambda \norm{\hat{\x} - \x}_1,
\end{equation}
where $\hat{\x}$ is the reconstruction and $\x$ is the ground truth image, after cropping to the central $320\times320$ region. Lambda was set to $0.001$. The models were trained for 20 epochs with a fixed learning rate of $0.0003$. All models were trained on a machine with 8 NVIDIA Volta V100 GPUs using data parallel training for about 3 days.

The U-Net models applied either to 15-coil k-space input or 15-coil image input start with 384 channels, which are doubled after each pooling. The GrappaNet model contains a total of 4 U-Nets, each of which starts with 192 channels. All three models have roughly 480M parameters.

Experimental results are shown in table \ref{tbl:results}, which lists three metrics that are computed in the same manner as \cite{zbontar2018fastmri}: normalized mean squared error (NMSE), peak signal to noise ratio (PSNR) and structural similarity (SSIM) \cite{wang2004image}. 
All of the proposed models perform significantly better than the baselines. The large difference in performance between a U-Net applied to all 15 coils versus the U-Net baseline underscores the importance of letting the neural network figure out how to combine the coil images.

The GrappaNet performs best according to all metrics. The improved performance of the GrappaNet can be attributed to the inclusion of the GRAPPA layer to implement parallel imaging within the network.

Some example reconstructions are shown in figures \ref{fig:example-4x} and \ref{fig:example-8x} for $4\times$ and $8\times$ accelerations, respectively. Figure \ref{fig:example-4x-crops} and \ref{fig:recons-arrows} show some of the medically relevant regions zoomed in for $4\times$ acceleration. The baseline U-Net model is able to remove aliasing artifacts, but this comes at the cost of severe over-smoothing. The reconstruction lacks some of the high frequency detail that is clinically relevant. The reconstructions from the image U-Net model are significantly better than the baseline, but they are not as sharp as the reconstructions from the GrappaNet model.

The Variational Net model makes heavy use of estimated sensitivity maps throughout the network, including in the data consistency terms. It is able to generate good reconstructions with $4\times$ acceleration, which retains a sufficient number of low frequency lines to estimate sensitivity maps. When the acquisition is accelerated by $8\times$, however, the performance degrades significantly since it is not possible to accurately estimate sensitivity maps for this case.

\section{Conclusion and Future Work} \label{sec:conclusions}

In this paper, we introduced the GrappaNet architecture for multi-coil MRI reconstruction. Multi-coil MRI reconstruction presents an important and challenging problem due to the prevalence of parallel imaging and the need to make scan-specific adaptations to the neural networks. GrappaNet addresses this challenge by integrating traditional parallel imaging methods with neural networks and training the model end-to-end. This allows the model to generate high fidelity reconstructions even at high acceleration factors.

The GRAPPA kernel used in the GrappaNet model is estimated from the low-frequency lines of k-space and is used as a fixed input to the model. A possible extension to this work could explore methods to optimize the process of estimating the kernel jointly with the rest of the network during training.

Quantitative measures such as NMSE, PSNR, and SSIM only provide an estimate for the quality of the reconstructions. Clinically important details are often subtle and contained in small portions of an MRI. Before techniques such as those presented in this paper can be used in practice, proper clinical validation studies need to be performed to ensure that the use of accelerated MRIs does not degrade the quality of diagnosis.

\clearpage

\bibliographystyle{plainnat}
\bibliography{grappa}

\clearpage

\appendix

\section{Dithering as post-processing}

The GrappaNet model was trained to optimize a linear combination of Structural Similarity \cite{wang2004image} and L1 loss between the reconstruction and the ground truth image. SSIM and L1 loss are imperfect proxies for radiologists' visual perception; optimizing SSIM, L1 loss, or a linear combination of them can produce unnaturally smooth reconstructions even when preserving diagnostic content. We can enhance the perceived sharpness of the images by adding low levels of noise, that is, by dithering. As established by~\cite{perlin1985image}, filtered noise (``Perlin noise'') is a good model for the synthesis of natural textures --- natural-looking textures include some noise. Quoting~\cite{pham2016noise-added}, ``the preservation of film grain noise can also help enhance the subjective perception of sharpness in images, known as acutance in photography, although it degrades the signal-to-noise ratio. The intentional inclusion of noise in processing digital audio, image, and video data is called dither.''

To avoid obscuring dark areas of the reconstruction by adding too much noise, we adapt the level of noise to the brightness of the image around each pixel. Specifically, we first normalize the image we wish to dither by dividing each pixel by the maximum pixel intensity in the image; then we blur the normalized image with a median filter taking medians over patches 11 pixels high by 11 pixel wide, then take the square root of the value at each pixel of the blurred image, and finally add to the image being dithered centered Gaussian noise of standard deviation $\sigma$ times the associated blurred pixel. We set $\sigma$ = 0.025 for non-fat-suppressed images and $\sigma$ = 0.05 for fat-suppressed images (which have a worse native SNR).

Examples of GrappaNet reconstructions with and without noise are shown in figures \ref{fig:grappa_noise-4x} and \ref{fig:grappa_noise-8x}. The dithered images look more natural, especially the PDFS images with $8\times$ under-sampling. The metrics reported in the main paper do not include this added noise.

\begin{figure*}
    \centering
    \includegraphics[page=1,width=\textwidth,keepaspectratio]{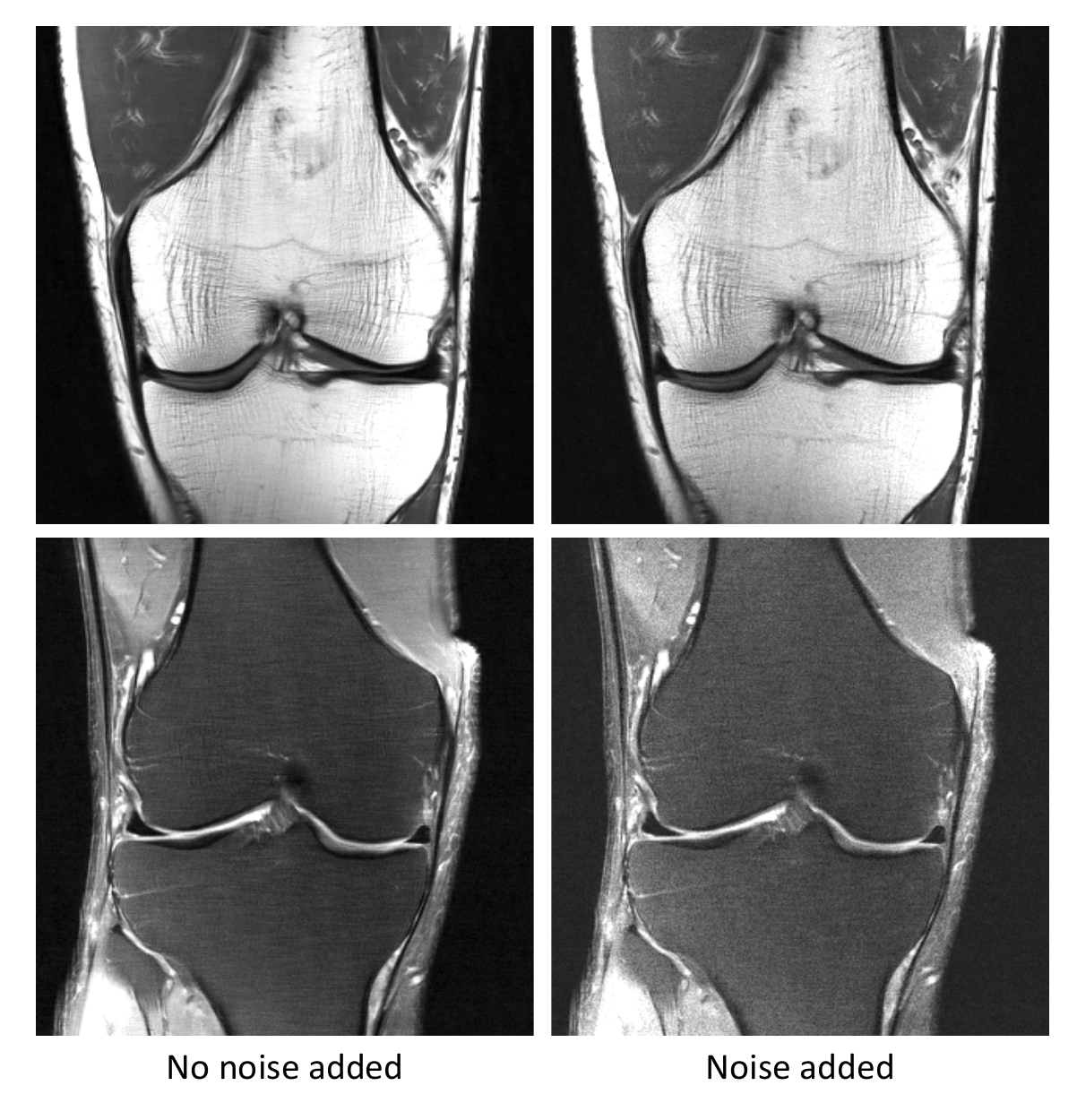}
    \caption{Example reconstructions from the GrappaNet model with $4\times$ under-sampling with and without dithering. The top row shows PD images without fat suppression and the bottom row shows PDFS images.}
    \label{fig:grappa_noise-4x}
\end{figure*}

\begin{figure*}
    \centering
    \includegraphics[page=2,width=\textwidth,keepaspectratio]{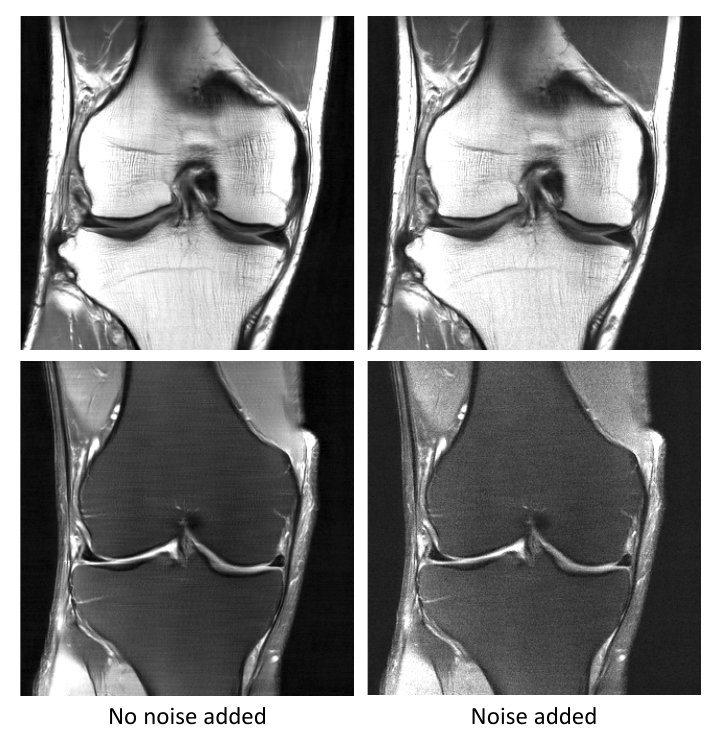}
    \caption{Example reconstructions from the GrappaNet model with $8\times$ under-sampling with and without dithering. The top row shows PD images without fat suppression and the bottom row shows PDFS images.}
    \label{fig:grappa_noise-8x}
\end{figure*}

\section{Training with random masks to counter adversarial examples}

Compressed sensing is the reconstruction of images to a resolution
beyond what reconstruction via classical signal processing would permit
for the amount of measurements actually made.
In MRI, the measurements are taken in $k$-space,
and the classical signal processing involves an inverse Fourier transform.
Compressed sensing reconstructs to the same resolution as if using
an inverse Fourier transform on more measurements than actually taken;
compressed sensing must be nonlinear to succeed.
When taking measurements in $k$-space at fixed locations,
it is relatively straightforward to construct objects whose measurements
at these fixed locations will result in reconstructions from compressed sensing
that are horribly wrong: simply alter arbitrarily the objects
in the parts of $k$-space in between those locations in $k$-space
that are actually measured. Whether such so-called ``adversarial'' examples
of objects being measured are worrisome depends
on where the actual measurements are made and (especially) on the algorithm
used for reconstruction.

If the algorithm used for reconstruction is trained on a set of examples
with measurements always taken at the same locations in $k$-space,
then the reconstruction is likely to be blind to properties of objects
that depend on parts of $k$-space in between those actually measured.
The adversarial examples can then hide horrible problems
in between the parts of $k$-space that are actually measured;
the algorithm for reconstruction trained on only fixed locations
in $k$-space will have no hope of learning how the unmeasured parts
of $k$-space contribute to the correct reconstruction.
On the contrary, if the algorithm used for reconstruction is trained
on examples with measurements taken at random locations in $k$-space
(which is particularly advantageous if each example gets measured
at several different random realizations of the sampling pattern),
and the random locations cover all $k$-space (over enough random realizations),
then the algorithm is likely to learn about all parts of $k$-space
during training
(here, ``all'' $k$-space refers to the sampling pattern used
for conventional reconstruction via the inverse Fourier transform
at full resolution).
When taking measurements at random locations in $k$-space,
the algorithm for reconstruction will probably detect at least a piece
of any adversarial attempt to hide horrible artifacts in parts
of $k$-space, and will learn how all relevant parts of $k$-space affect
the correct reconstruction.

Therefore, a machine-learned algorithm for reconstruction should train
on examples measured at randomized locations in $k$-space
in order to avoid some adversarial examples,
such as those constructed by~\cite{antun-renna-poon-adcock-hansen}.
Moreover, the measurements for the validation and testing sets
must also be randomized, in the following subtle sense:
the locations of the measurements in $k$-space
must be stochastically independent of the object being imaged.
Ideally the object will be deterministic and the locations of the measurements
in $k$-space will be drawn randomly independently of the object.
Thus, the object being imaged should not be constructed conditional
on knowing the locations in $k$-space of the measurements being taken;
an object in physical reality has no way of knowing where the measurements
are being taken.
The adversarial examples of~\cite{antun-renna-poon-adcock-hansen}
construct objects that depend on where the measurements are being taken,
and so are inapplicable to the setting of randomized locations
for the measurements.
In practice, the same random locations in $k$-space can be used
for multiple objects, provided that the objects being imaged cannot alter
themselves based on knowing where the measurements are being taken,
and provided that the training of any machine-learned reconstruction
considers many different random locations in $k$-space
(preferably covering all $k$-space over enough random realizations).

To summarize:
\begin{enumerate}
\item Measurements should be at randomized locations in $k$-space
during training of machine-learned algorithms for reconstruction,
such that the random locations cover all $k$-space
(over enough random realizations), where ``all'' refers
to the sampling pattern used for conventional reconstruction
at full resolution via the inverse Fourier transform.
\item Measuring each object in the training set
at multiple different random samples in $k$-space is ideal,
constituting a kind of data augmentation that regularizes the reconstruction
and improves generalization and robustness to adversarial examples.
\item The object being imaged in reality during validation and testing
should be deterministic, with the random locations in $k$-space
where measurements are taken being stochastically independent of the object.
\item When taking measurements at randomized locations in $k$-space,
the object should not alter itself based on where measurements are made;
adversarial examples are irrelevant when they are conditional
on knowing the locations of the randomized measurements.
\item The same random locations in $k$-space can be used across the objects
in the validation and testing sets
(yet these locations must vary during training!).
\end{enumerate}

Fortuitously, algorithms for reconstruction that obey the above conditions
are also ideal for use in estimating errors via the bootstrap,
as described by~\cite{defazio-tygert-ward-zbontar}.

Regarding technologically reasonable sampling patterns,
MRI works well taking measurements along the following lines:
\begin{enumerate}
\item radial lines in $k$-space, with the lines at random angles
\item parallel lines in $k$-space, with the lines at random offsets
\item equispaced parallel lines in $k$-space,
with the overall offset chosen at random
\end{enumerate}
In all cases, ``random'' means the same angle or offset for different objects
being imaged in validation and testing sets,
but with the angles or offsets varying at random during training
and in the bootstrap or jackknife error estimation.
So-called parallel imaging usually supplements the above measurements
with some additional measurements of mainly low frequencies for autocalibration
of sensitivity maps or of convolutional kernels for fusing contributions
from multiple receiver coils,
as discussed by~\cite{antun-renna-poon-adcock-hansen} and others.
The extra set of autocalibration measurements is merely a bonus, not requiring
the same randomization as the other measurements.

\section{Example Reconstructions}
Additional example reconstructions picked at random from the validation set are shown in figures \ref{fig:pd4x}-\ref{fig:pdfs8x}. In each case, the images on the left are the ground truths and the images on the right are the dithered reconstructions.

\begin{figure*}
    \centering
    \includegraphics[page=1,keepaspectratio]{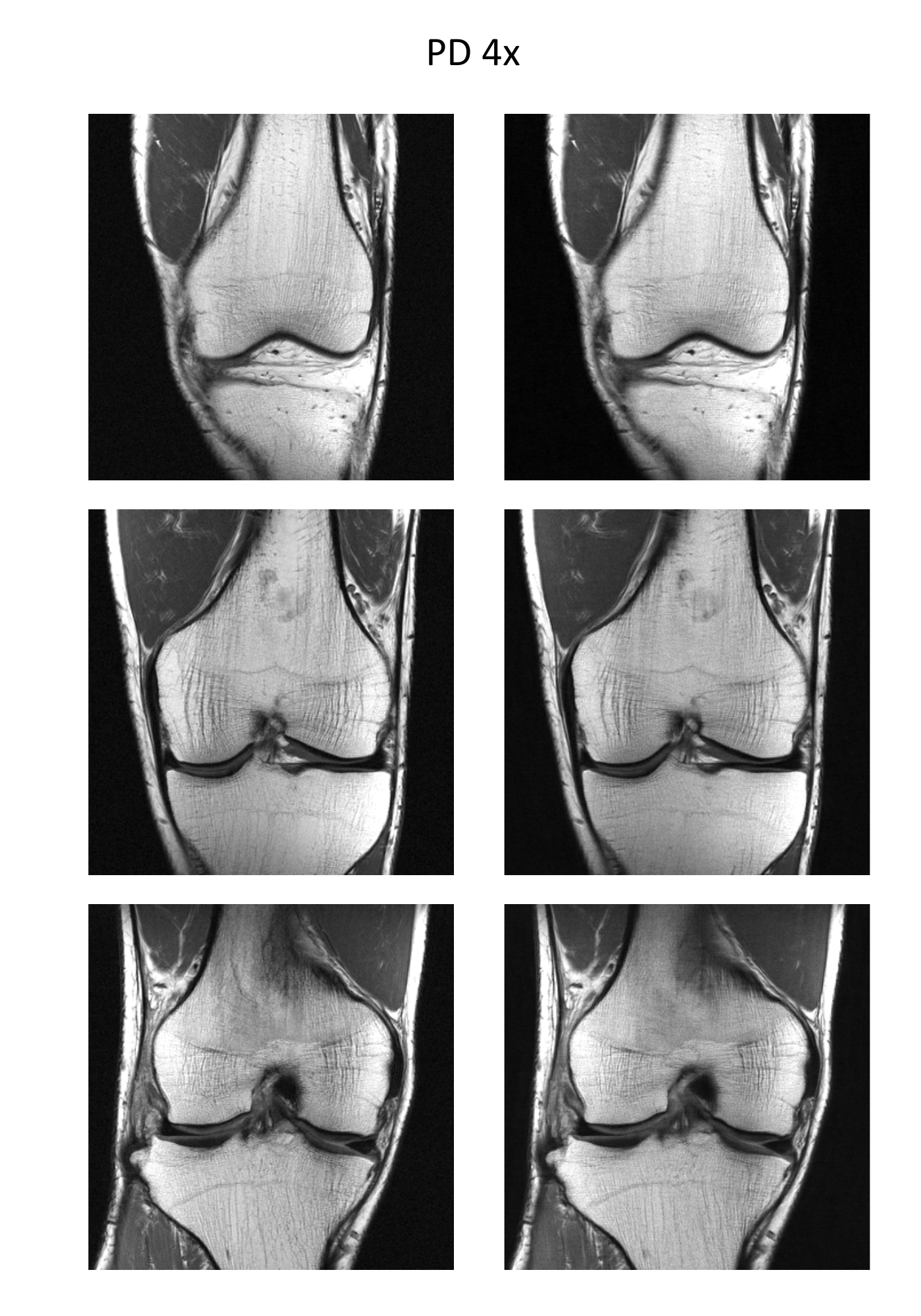}
    \caption{Proton Density with $4\times$ under-sampling.}
    \label{fig:pd4x}
\end{figure*}

\begin{figure*}
    \centering
    \includegraphics[page=2,keepaspectratio]{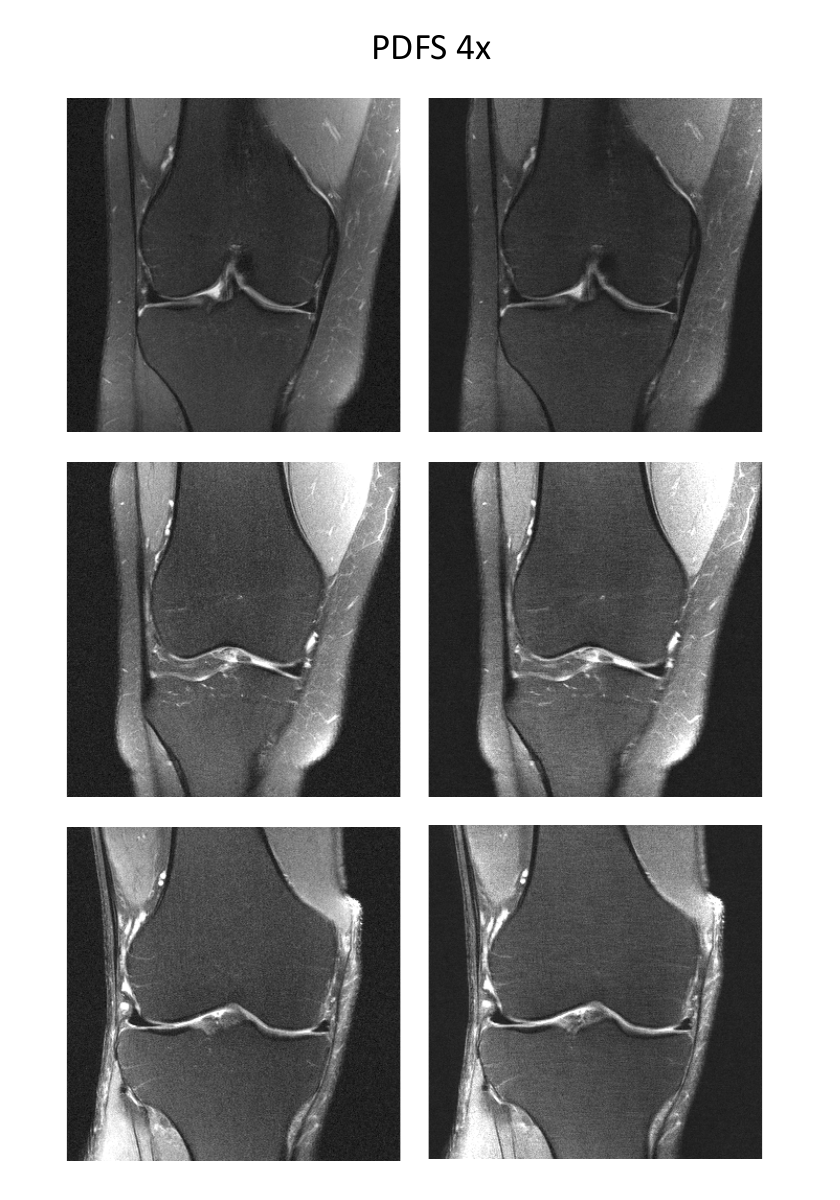}
    \caption{Proton Density with Fat Suppression with $4\times$ under-sampling.}
    \label{fig:pdfs4x}
\end{figure*}

\begin{figure*}
    \centering
    \includegraphics[page=3,keepaspectratio]{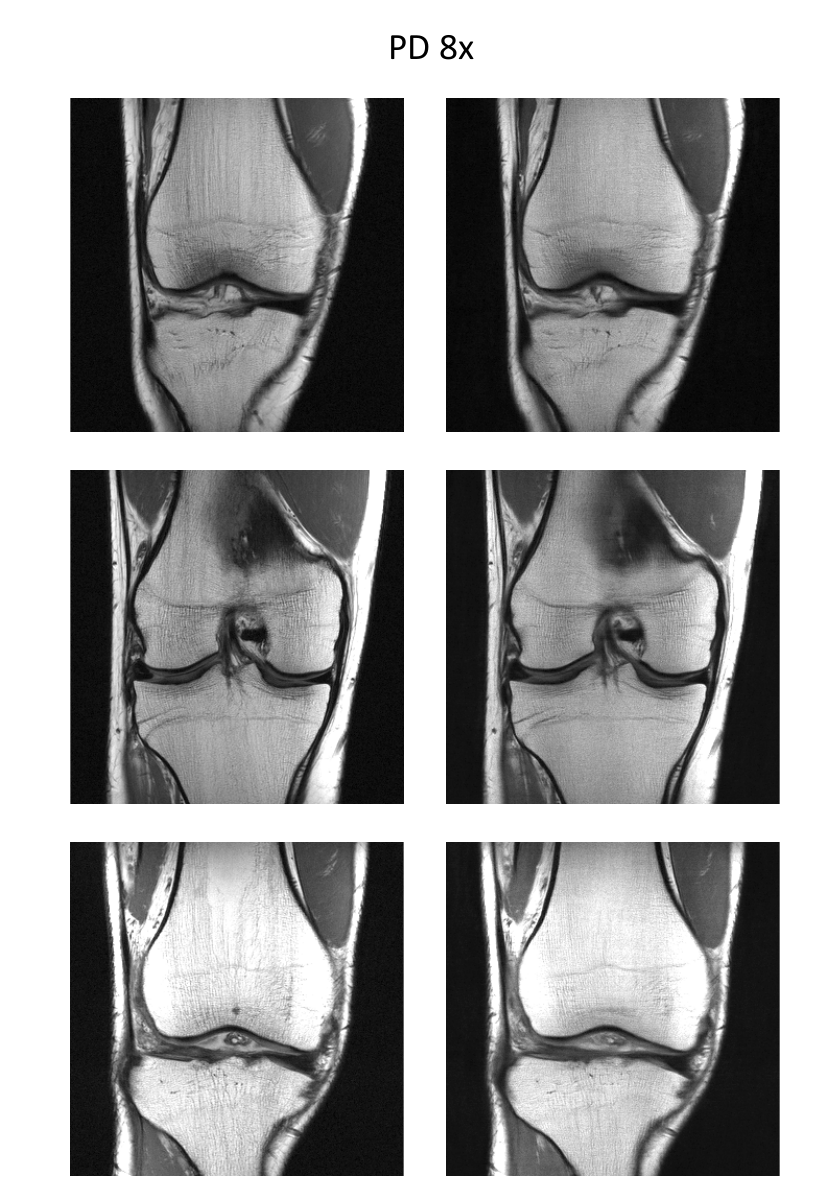}
    \caption{Proton Density with $8\times$ under-sampling.}
    \label{fig:pd8x}
\end{figure*}

\begin{figure*}
    \centering
    \includegraphics[page=4,keepaspectratio]{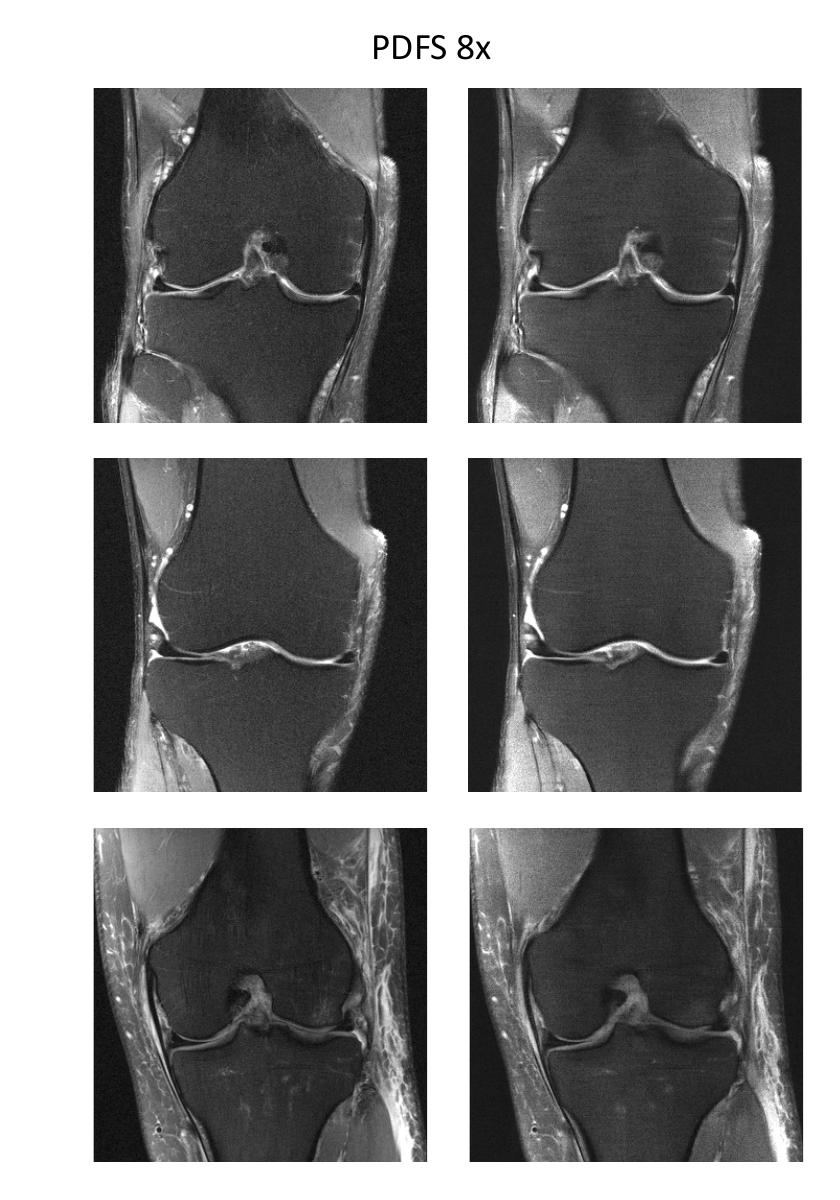}
    \caption{Proton Density with Fat Suppression with $8\times$ under-sampling.}
    \label{fig:pdfs8x}
\end{figure*}


\end{document}